\documentclass[pdflatex,sn-mathphys-num]{sn-jnl}

\usepackage{graphicx}%
\usepackage{multirow}%
\usepackage{amsmath,amssymb,amsfonts}%
\usepackage{amsthm}%
\usepackage{mathrsfs}%
\usepackage[title]{appendix}%
\usepackage{xcolor}%
\usepackage{textcomp}%
\usepackage{manyfoot}%
\usepackage{booktabs}%
\usepackage{algorithm}%
\usepackage{algorithmicx}%
\usepackage{algpseudocode}%
\usepackage{listings}%
\usepackage{comment}
\usepackage{bm}
\usepackage{physics}
\theoremstyle{thmstyleone}%
\theoremstyle{thmstyletwo}%
\theoremstyle{thmstylethree}%

\begin{document}

\title[Article Title]{Necessary conditions for spin-resolved Josephson diode effect across strongly spin-polarized magnetic materials}

\author*[1]{\fnm{Danilo} \sur{Nikoli\'c}}\email{danilo.nikolic@uni-greifswald.de}

\author[1]{\fnm{Niklas L.} \sur{Schulz}}\email{niklas.schulz@uni-greifswald.de}

\author[1]{\fnm{Matthias} \sur{Eschrig}}\email{matthias.eschrig@uni-greifswald.de}

\affil[1]{\orgdiv{Institute of Physics}, \orgname{University of Greifswald}, \orgaddress{\street{Felix-Hausdorff-Str. 6}, \postcode{17489}, \city{Greifswald}, \country{Germany}}}

\abstract{We present a set of necessary conditions for the appearance of charge and spin Josephson diode effects across strongly spin-polarized inhomogeneous magnetic materials (FM) placed between two spin-singlet superconductors. Noncoplanarity of the FM's spin texture gives rise to quantum geometric phases, $\Delta\varphi'$, that enter the Josephson current-phase relation (CPR) similarly to the superconducting phase difference, resulting in charge and spin Josephson diode effects. Our study shows that such effects appear if the CPR possesses no phase-inversion center, achieved under the following conditions. First, noncoplanarity of the spin texture is necessary to break the spatial inversion symmetry. Second, both spin bands have to contribute to the transport, i.e., the effect is absent in half-metallic junctions. Third, different band-specific densities of states are required, and this condition is ensured by the strong spin polarization of the FM. Finally, higher harmonics in the CPR are necessary, i.e., the effect is absent in the tunneling limit. However, even in this case, the CPR must not have a phase-inversion center, which is ensured by the restriction of the quantum geometric phase to values $\Delta\varphi'\neq k\pi/2, k\in\mathbb{Z}$. We formulate a minimal phenomenological model that incorporates all these points, qualitatively illustrating our theory.}




\maketitle
\section{Introduction}\label{sec1}
The current-phase relation (CPR) of a Josephson junction is a $2\pi$-periodic function expressing the Josephson current $I$ as a function of the superconducting phase difference $\Delta \chi$, i.e. $I(\Delta\chi)=I(\Delta\chi+2\pi)$. Under time reversal, the current $I$ as well as the superconducting phase $\chi $ reverse sign, i.e. $I(\Delta\chi)\to -I(-\Delta\chi )$. Under inversion, on the other hand, the current $I$ reverses sign, and the superconducting phase $\chi(\vec{r})\to \chi(-\vec{r})$, i.e. again $I(\Delta\chi)\to -I(-\Delta\chi )$.
If the Josephson junction is invariant under either time reversal or inversion, i.e. $I(\Delta\chi)= -I(-\Delta\chi )$, then the current-phase relation is an odd function of the superconducting phase difference, and it follows
$I(\Delta\chi=0)=0$. Furthermore, as $2\pi$-periodicity requires $I(-\pi)=I(\pi)$, it follows also that $I(\Delta\chi=\pi)=0$, and therefore $I(n\pi)=0, n\in\mathbb{Z}$, 
so that consideration of the interval $0\leq\Delta\chi\leq\pi$ is typically sufficient. 
This effect became known as the normal Josephson effect~\cite{josephsonPossibleNewEffects1962}, and there are some general properties of it, irrespective of the particular system under study~\cite{golubovCurrentphaseRelationJosephson2004}. The dc supercurrent is only possible if there is a superconducting phase gradient. 
In general, the Josephson CPR is nonsinusoidal, and the following harmonic decomposition can be performed:
\begin{equation}\label{eqn:CPR_general}
    I(\Delta\chi) = \sum_{n\geq 0}\qty[I_n\sin(n\Delta\chi)+J_n\cos(n\Delta\chi)],
\end{equation}
where the $n$ index refers to the process involving the coherent transfer of $n$ Cooper pairs. 
The term with $n=0$ is only present if several transport channels (e.g. spin channels) can carry equal current in opposite direction.
If the junction possesses time-reversal or inversion symmetry, then $J_n=0, \forall n\in\mathbb{Z}$ leading to the normal Josephson effect. 
Note that it does not follow from invariance under time-reversal or inversion that the equilibrium phase has to be either 0 or $\pi $, as additional phase difference values $\pm \chi$ with zero current are in general possible, and under certain circumstances some of those can minimize the junction energy \cite{Yip1995,goldobinJosephsonJunctionsSecond2007}.
In such a case the junction is called a $\chi$-junction~\cite{buzdinPeriodicAlternating02003}, and the presence of time reversal or inversion symmetry leads to pairwise degenerate junction ground states. A $\chi$-junction requires higher harmonics. In the tunnelling limit, where the leading term $I_1\sin (\Delta \chi)$ dominates, a 0 or $\pi $ junction is realized depending on the sign of $I_1$.
In any case, whenever $I(\Delta\chi)= -I(-\Delta\chi )$ holds
the critical Josephson currents in positive and negative current direction are equal, i.e. the transport is reciprocal. 

If time-reversal and inversion symmetries are broken, the anomalous Josephson effect may appear, $I(\Delta\chi)\neq -I(-\Delta\chi)$. Here, we can distinguish two regimes. If the Josephson CPR possesses a \textit{phase-inversion center} $\chi_0$ around which the supercurrent is antisymmetric, $I(\Delta\chi+\chi_0)=-I(-\Delta\chi+\chi_0)$,   the CPR is of the form
\begin{equation}
	I(\Delta\chi) = \sum_{n\geq 1} \tilde{I}_n\sin[n(\Delta\chi-\chi_0)],
\end{equation}
which can be related to Eq.~\eqref{eqn:CPR_general} via $I_n=\tilde{I}_n\cos\chi_0$, $J_n=-\tilde{I}_n\sin\chi_0$ for $n\ge 1$, and $J_0=0$.
It is easy to see that in this case transport is reciprocal, despite the fact that higher harmonics are present and $I(\Delta\chi)\ne -I(-\Delta\chi )$. Such a realization is known as a "$\chi_0$-junction", and this effect has been predicted in various setups such as junctions with unconventional superconductors~\cite{GeshkenbeinLarkin1986,Yip1995,Sigrist1998}, strong spin-orbit coupling~\cite{buzdinDirectCouplingMagnetism2008}, or strongly spin-polarized itinerant ferromagnets~\cite{eschrigTripletSupercurrentsClean2008}; some of which have been experimentally verified~\cite{szombatiJosephsonPh0junctionNanowire2016,reinhardtLinkSupercurrentDiode2024}. However, if the Josephson CPR has no phase-inversion center, 
 the critical current in one direction $(+)$ in general differs from that in the opposite direction $(-)$, leading to the so-called superconducting or Josephson diode effect (JDE), which recently has been studied in numerous experimental \cite{andoObservationSuperconductingDiode2020, baumgartnerSupercurrentRectificationMagnetochiral2022, costaSignReversalJosephson2023, gutfreundDirectObservationSuperconducting2023, houUbiquitousSuperconductingDiode2023, nadeemSuperconductingDiodeEffect2023, strambiniSuperconductingSpintronicTunnel2022, trahmsDiodeEffectJosephson2023} and theoretical works \cite{greinSpinDependentCooperPair2009, margarisZeroPhaseDifference2010,misakiTheoryNonreciprocalJosephson2021,daidoIntrinsicSuperconductingDiode2022, fominovAsymmetricHigherharmonicSQUID2022, haltermanSupercurrentDiodeEffect2022, hePhenomenologicalTheorySuperconductor2022, ilicTheorySupercurrentDiode2022, tanakaTheoryGiantDiode2022, yuanSupercurrentDiodeEffect2022, zhangGeneralTheoryJosephson2022, zinklSymmetryConditionsSuperconducting2022,soutoJosephsonDiodeEffect2022,steinerDiodeEffectsCurrentBiased2023,costaMicroscopicStudyJosephson2023,costaSignReversalJosephson2023,kopasovAdiabaticPhasePumping2023,meyerJosephsonDiodeEffect2024,putilovNonreciprocalElectronTransport2024,sunVoltagetunableSpinSupercurrent2025a,patilSpinsplitAndreevBound2025}. Typically, this effect is quantified by the so-called diode efficiency, defined as 
\begin{equation}
\eta=\frac{I^+-|I^-|}{I^+ + |I^-|}
\end{equation}
where $I^\pm$ refers to the critical current in the corresponding direction.  

In this paper, we establish a set of necessary conditions for the appearance of the JDE in junctions with strongly spin-polarized materials. 
In junctions containing a magnetized material, characterized by an exchange field $\vec{J}$, the behavior under time reversal leads to the relation $I(\Delta\chi,\vec{J})=-I(-\Delta\chi,-\vec{J})$.
This follows from the fact that the Hamiltonian term $H_1(\vec{J})=-\vec{J}\cdot\vec{\sigma}$ transforms under time reversal into $H_1(-\vec{J})$.
For a homogeneous exchange field and singlet superconducting pair potential, the Josephson charge current does not depend on the direction but only on the magnitude of $\vec{J}$, such that no JDE appears. If an inhomogeneous spin texture is present, characterized by an inhomogeneous exchange field $\vec{J}(\vec{r})$, then the Josephson charge current may depend also on the direction of $\vec{J}$. 
Then, $I[\Delta\chi,\vec{J}(\vec{r})]=-I[-\Delta\chi,-\vec{J}(\vec{r})]$.
Inversion transforms $H_1[\vec{J}(\vec{r})]$ into $H_1[\vec{J}(-\vec{r})]$, such that
$I[\Delta\chi,\vec{J}(\vec{r})]=-I[-\Delta\chi,\vec{J}(-\vec{r})]$ follows.\footnote{More generally, for a $\vec{k}$-dependent exchange field, which in addition can vary spatially on a scale much larger than the Fermi wavelength, the time reversal involves $-\vec{J}(-\vec{k},\vec{r})$, and inversion involves $\vec{J}(-\vec{k},-\vec{r})$.} 

For example, let us assume a ferromagnetic trilayer characterized by $\vec{J}_L$, $\vec{J}$, and $\vec{J}_R$,
and let us assume that the Josephson charge current depends on the nonlocal quantity $\eta\equiv (\vec{J}_L\times \vec{J}_R)\cdot \vec{J}$. Reversing all three exchange fields leads to $-\eta $. Likewise, inversion leads to $(\vec{J}_R\times \vec{J}_L)\cdot \vec{J}=-\eta $. Thus, in this case, $I(\Delta\chi,\eta )=-I(-\Delta\chi,-\eta )$ holds. Consequently, if $\eta $ is nonzero, then in general $I(\Delta\chi,\eta )\ne -I(-\Delta\chi,\eta )$.  In such a case, a JDE may occur provided a phase-inversion center is absent.
Note, that $\eta $ is non-zero only for a noncoplanar arrangement of the three exchange fields.

In strongly spin-polarized ferromagnetic junctions, Josephson transport is entirely mediated by long-range equal-spin triplet correlations. If, additionally, such materials display noncoplanar spin textures, the Josephson diode effect appears due to so-called quantum geometric phases~\cite{greinSpinDependentCooperPair2009}. These phases are nonlocal and are induced by a noncoplanar magnetization profile in a ferromagnetic trilayer~\cite{schulzQuantumgeometricSpinCharge2025,schulzTheoryQuantumgeometricCharge2025a} or by intrinsically noncoplanar magnetization texture, such as in a conical magnet~\cite{nikolicSpinresolvedJosephsonDiode2025}. In what follows, we discuss their physical origin in more detail. 
\begin{figure}
    \centering
    \includegraphics[width=1\linewidth]{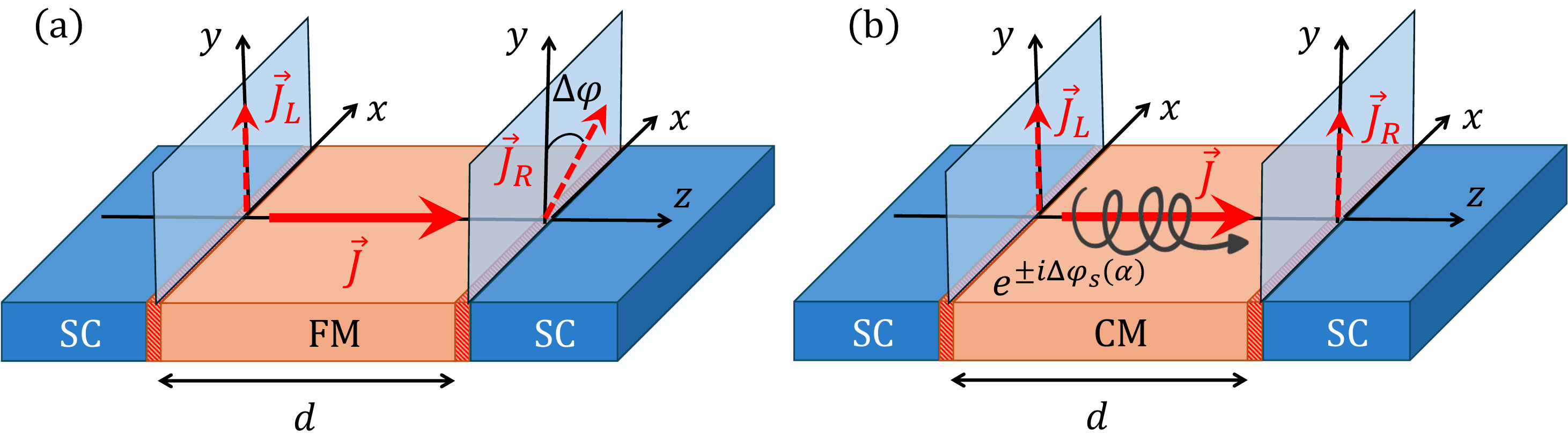}
    \caption{Two configurations of strongly spin-polarized magnetic materials that exhibit quantum geometric phases. Panel (a) shows a ferromagnetic trilayer where the exchange field of the central metallic layer (FM), $\Vec{J}$, sets a global quantization axis and the quantum geometric phase $\Delta\varphi$ is determined by the relative azimuthal angle between the exchange fields of the satellite ferromagnetic layers, $\vec{J}_L$ and $\vec{J}_R$. Panel (b) shows a conical magnet (CM) that in the so-called adiabatic approximation (see text) can be viewed as a ferromagnet with the opposite spin geometric phase $\Delta\varphi_s$ of the two spin bands.}
    \label{fig:system}
\end{figure}
\section{Geometric phases in strongly spin-polarized materials}
To show the origin of quantum geometric phases appearing in ferromagnetic trilayers, we consider an interface between a BCS superconductor (SC) and a uniform metallic ferromagnet FM connected via a thin ferromagnetic insulating layer (FI). Considering the FM's exchange field $\vec{J}$ as a global spin-quantization axis (therefore, along z direction), let us examine the spin rotation onto the FI's exchange field $\vec{J}_L$ pointing in an arbitrary direction $\vec{n}_L=(\sin\alpha_L\cos\varphi_L,\sin\alpha_L\sin\varphi_L,\cos\alpha_L)$:
 \begin{equation}
    \begin{pmatrix} \uparrow \\ \downarrow \end{pmatrix}_{\vec{n}_L} = \begin{pmatrix} \cos\frac{\alpha_L}{2}  &  \sin{\frac{\alpha_L}{2}e^{i\varphi_L}} \\ -\sin{\frac{\alpha_L}{2}e^{-i\varphi_L}}  & \cos\frac{\alpha_L}{2} \end{pmatrix} \begin{pmatrix} \uparrow \\ \downarrow \end{pmatrix}_z.
 \end{equation}
 Consequently, the pair amplitudes from the SC transform as follows:
 \begin{align}
 \label{eqn:singlet}
     (\uparrow\downarrow-\downarrow\uparrow)_{\vec{n}_L} &=   (\uparrow\downarrow-\downarrow\uparrow)_z,\\
\label{eqn:triplet}
     (\uparrow\downarrow+\downarrow\uparrow)_{\vec{n}_L} &= -\sin\alpha_L\left[e^{-i\varphi_L}(\uparrow\uparrow)_z-e^{i\varphi_L}(\downarrow\downarrow)_z\right] +\cos\alpha_L (\uparrow\downarrow+\downarrow\uparrow)_z.
 \end{align}
 Therefore, once mixed-spin triplet correlations [see Eq.~\eqref{eqn:triplet}] are formed along $\vec{J}_L$, they give rise to equal-spin triplet correlations along $\vec{J}$ in the FM. Furthermore, the two equal-spin triplet amplitudes, $(\uparrow\uparrow)_z$ and $(\downarrow\downarrow)_z$, acquire a relative phase $\pm(2\varphi_L+\pi)$ with respect to each other. Considering another spin rotation at the right SC/FM interface featuring another thin ferromagnetic insulator, characterized by $\vec{J}_R=J_R(\sin\alpha_R\cos\varphi_R,\sin\alpha_R\sin\varphi_R,\cos\alpha_R)$, gives rise to the \textit{quantum geometric phase difference} determined by the relative phase shift at the two interfaces, $\Delta\varphi=\varphi_R-\varphi_L$. A scheme of a system under study is shown in Fig.~\ref{fig:system}(a), where, for simplicity, we set $\alpha_L=\alpha_R=\pi/2$, which qualitatively does not alter the main conclusions.

 Note that the above considerations do not require the outer ferromagnetic layers of the trilayer to be insulating. They could also be metallic as long as their thickness is sufficiently small for all three triplet components to penetrate appreciably. The important issue is the misalignment of the exchange fields, which leads to triplet rotation and allows equal-spin pairs to enter the central ferromagnet~\cite{houzetLongRangeTriplet2007}. This mechanism works for both metallic and insulating ferromagnetic interface layers.

An analogous picture for intrinsically noncoplanar magnetic materials can be obtained considering a single electron interacting with a localized spin via the exchange interaction
\begin{equation}
\label{eqn:F_Hamiltonian}
\hat{H} = \frac{\hat{\vec{p}}\,^2}{2m} - J\vec{m}(\vec{r})\cdot\hat{\vec{{\sigma}}},
\end{equation}
where $\vec{m}(\vec{r})=[\sin\alpha(\vec{r})\cos\phi(\vec{r}),\sin\alpha(\vec{r})\sin\phi(\vec{r}),\cos\alpha(\vec{r})]$ is the spin texture of the material. The spin part of the Hamiltonian above can be diagonalized by the following SU(2) gauge transformation:
\begin{equation}
    \hat{U}(\vec{r})(J\vec{m}\cdot\hat{\vec{\sigma}})\hat{U}^\dagger(\vec{r}) = J\hat{\sigma}_3,
\end{equation}
where 
\begin{equation}
\label{eqn:U}
    \hat{U}(\vec{r}) = \exp[-i\frac{\alpha(\vec{r})}{2}\vec{n}(\vec{r})\cdot{\hat{\vec{\sigma}}}],
\end{equation}
with $\vec{n}=(\vec{m}\times \vec{e}_3)/|\vec{m}\times \vec{e}_3|=\sin{\left[\phi(\vec{r})\right]}\vec{e}_1-\cos{\left[\phi(\vec{r})\right]}\vec{e}_2$. Assuming slow spatial variations of the magnetic texture (adiabatic condition), the overlap between two states on the Bloch sphere can be evaluated as $\braket{\vec{m}',\pm}{\vec{m},\pm}=\mel{\uparrow\!/\!\downarrow}{\hat{U}(\vec{r}\,')\hat{U}^\dagger(\vec{r})}{\uparrow\!/\!\downarrow}\approx 1 \pm i\delta\varphi_s\approx e^{\pm i\delta\varphi_s}$, where 
 \begin{equation}
 \delta\varphi_s=\sin^2\left(\frac{\alpha}{2}\right)[\phi(\vec{r})-\phi(\vec{r}\,^\prime)]\approx \sin^2\left(\frac{\alpha}{2}\right)(\delta\vec{r}\cdot\vec{\nabla}\phi)
 \end{equation}
 is the so-called \textit{(adiabatic) spin gauge phase} with $\delta\vec{r}=\vec{r}-\vec{r}\,'$ being the relative coordinate. For a conical magnet, where $\alpha=$ const., the above phase can be integrated from $\vec{r}_1$ to $\vec{r}_2$, leading to a {\it spin geometric phase} 
 \begin{equation}\label{eqn:phi_s}
 \Delta \varphi_{s}=\sin^2\left(\frac{\alpha}{2}\right) \left[\phi(\vec{r}_2)-\phi(\vec{r}_1)\right].
 \end{equation}
Note that the nonlocality of Cooper pairs, which extend over mesoscopic distances, makes the geometric phase $\Delta\varphi_s$ \textit{nonlocal} on atomic length scales. Then, similarly to the quantum geometric phase introduced previously, the equal-spin pairing correlations with the opposite spins acquire the opposite spin geometric phases. Finally, integrating $\delta\varphi_s$ from $\varphi(\vec{r}_1)=0$ to $\varphi(\vec{r}_2)=2\pi$ yields $\Delta\varphi_s=2\pi\sin^2(\alpha/2)$, which corresponds to one half of the solid angle of the corresponding cone. Namely, $\Delta\varphi_s=\Omega/2$, where $\Omega=4\pi\sin^2(\alpha/2)$, and this gives a geometric interpretation of $\Delta\varphi_s$. A schematic view of the system in the adiabatic approximation is shown in Fig.~\ref{fig:system}(b).

\section{Charge and spin Josephson CPRs}
In strongly spin-polarized magnetic materials (FM) coupled to singlet superconductors (SC), mixed-spin correlations (spin singlets and triplets with $s_z=0$) are nonvanishing only in an atomically thin layer close to the FM/SC interface. In other words, they vanish over mesoscopic scales set by the superconducting coherence length, and the Josephson current is fully mediated by equal-spin triplet correlations (triplets with $s_z=\pm 1$), if the system allows for these. For this reason, we need magnetic inhomogeneity (either in a ferromagnetic multilayers or an intrinsically inhomogeneous magnetic materials such as conical magnets), and such a case allows for an insight into spin-resolved quantities, e.g., $\uparrow\uparrow$- and $\downarrow\downarrow$-supercurrents.

In strongly spin-polarized systems that feature geometric phases, spin-resolved currents transform under time reversal as
\begin{equation}
    I_{\sigma\sigma}(-\Delta\chi,-\Delta\varphi') = - I_{\sigma\sigma}(\Delta\chi,\Delta\varphi'),
\end{equation}
where $\sigma=\uparrow\downarrow$. This follows from a consideration of the Hamiltonian under time reversal, considering that each exchange field $\vec{J}$ originates from a mean-field treatment of spin Heisenberg interactions and consequently reverses its direction under time reversal. The spin quantization axis is given by the unit vector in direction of $\vec{J}$ of the local exchange field in the metallic ferromagnet, and therefore also reverses under time reversal (keeping the assignemnt between majority/minority spin and spin-up/spin down intact).
Numerical analysis of CPRs in such systems suggests the following harmonic expansion~\cite{schulzQuantumgeometricSpinCharge2025,schulzTheoryQuantumgeometricCharge2025a,nikolicSpinresolvedJosephsonDiode2025}:
\begin{align}
    I_{\uparrow\uparrow} &= \frac{1}{2}\sum_{\mu,\nu=-\infty}^\infty  (-1)^{\mu+\nu} \mu I_{\mu,\nu} \sin\psi_{\mu,\nu}, \label{eqn:I_uu} \\ 
    I_{\downarrow\downarrow} &= \frac{1}{2}\sum_{\mu,\nu=-\infty}^\infty  (-1)^{\mu+\nu} \nu I_{\mu,\nu} \sin\psi_{\mu,\nu}, \label{eqn:I_dd}
\end{align}
where $I_{-\mu,-\nu}=I_{\mu,\nu}$, and 
\begin{equation}\label{eqn:effJos}
\psi_{\mu,\nu} = (\mu+\nu)\Delta\chi - (\mu-\nu)\Delta\varphi^\prime
\end{equation}
is the \textit{effective Josephson phase}. The $I_{\mu,\nu}$ coefficients in the above expansion refer to the amplitudes of processes that involve transport $\mu$ $\uparrow\uparrow$-pairs and $\nu \downarrow\downarrow$-pairs across the junction. The sign of $\mu$ and $\nu$ refers to the direction of propagation of the corresponding pair. As introduced above, $\Delta\varphi'$ is the quantum geometric phase, which can be $\Delta\varphi$ or $\Delta\varphi_s$, but also a sum of the two, if the system under study allows for both (see Ref.~\cite{nikolicSpinresolvedJosephsonDiode2025}). Note that these phases enter the Josephson CPRs very similarly to the superconducting phase difference, $\Delta\chi$, due to their U(1) nature. The factor $(-1)^{\mu+\nu}$ in Eqs.~\eqref{eqn:I_uu} and~\eqref{eqn:I_dd} reflects the fact that in strongly spin-polarized materials the ground state is a $\pi$-junction. However, this fact does not change the physics of the JDE and, henceforth, we omit this factor, i.e., we assume that the model junction is in a $0$-state in the absence of any geometric phase. Having obtained the spin-resolved currents, the charge and spin currents can be simply computed as, respectively,
\begin{equation}
    I_\mathrm{ch} = 2e(I_{\uparrow\uparrow}+I_{\downarrow\downarrow}) \quad\text{and}\quad I_\mathrm{sp} = \hbar(I_{\uparrow\uparrow}-I_{\downarrow\downarrow}). 
\end{equation}

An alternative way for obtaining the above expression is by introducing the Josephson energy $E_J$ (again, we omit the $(-1)^{\mu+\nu}$ factor)~\cite{schulzQuantumgeometricSpinCharge2025}:
\begin{equation}\label{JE}
    E_J(\Delta \chi, \Delta \varphi') =-\frac{\hbar}{2} \sum_{\mu,\nu=-\infty}^\infty I_{\mu,\nu}\cos \psi_{\mu,\nu},
\end{equation}
which implies
\begin{align}
     \label{eqn:I_ch}
     I_{\rm ch}(\Delta \chi, \Delta \varphi')&=2e(I_{\uparrow\uparrow}+I_{\downarrow\downarrow})=
    \frac{2e}{\hbar}\frac{\partial E_J}{\partial \Delta \chi}=e\sum_{\mu,\nu=-\infty}^\infty (\mu+\nu)I_{\mu,\nu}\sin\psi_{\mu,\nu},\\
    \label{eqn:I_sp}
    I_{\rm sp }(\Delta \chi, \Delta \varphi')&=2S (I_{\uparrow\uparrow}-I_{\downarrow\downarrow})= -\frac{2S}{\hbar}\frac{\partial E_J}{\partial \Delta \varphi'} = S\sum_{\mu,\nu=-\infty}^\infty (\mu-\nu)I_{\mu,\nu}\sin\psi_{\mu,\nu}.
\end{align}
Here, $S=\frac{\hbar}{2}$, and $e<0$ denotes the elementary charge. In the following, Eqs.~\eqref{JE}-\eqref{eqn:I_sp} will be used to formulate the necessary conditions for the appearance of charge and spin Josephson diode effects in junctions featuring quantum geometric phases.

\section{Necessary conditions for JDE}\label{sec:Results}
As already anticipated in the introduction, our aim here is to formulate a set of rules that apply to junctions involving strongly spin-polarized materials to exhibit the Josephson diode effect. From the practical point of view, we examine the conditions under which the CRPs given in the preceding section \textit{do not} possess a phase-inversion center, namely, there is no $\chi_0$ such that $I(\Delta\chi+\chi_0)=-I(-\Delta\chi+\chi_0)$. Since the Josephson charge current is a derivative of the free energy with respect to $\Delta\chi$ [see Eq.~\eqref{eqn:I_ch}], the latter condition implies $E_J(\Delta\chi+\chi_0)=+E_J(-\Delta\chi+\chi_0)$. In what follows, we investigate under which conditions this relation holds or gets violated. 
\subsection{Inversion-symmetric magnetic spin texture}
Let us first discuss the role of the noncoplanarity of the magnetic texture. Three vectors are noncoplanar if and only if their triple product is nonzero. Considering a ferromagnetic trilayer depicted in Fig.~\ref{fig:system}(a), this condition translates as$(\vec{J}_L\times\vec{J}_R)\cdot\vec{J}=|\vec{J}_L||\vec{J}_R||\vec{J}|\sin\alpha_L\sin\alpha_R\sin(\varphi_R-\varphi_L)\neq 0$ which implies $\varphi_R-\varphi_L=\Delta\varphi\neq0$ (see also Sec.~\ref{sec1}). Therefore, since we already set $\alpha_L=\alpha_R=\pi/2$, the noncoplanarity condition is fully determined by the quantum geometric phase $\Delta\varphi$. Considering a conical magnet, the noncoplanarity condition can be written as $[\vec{m}(x)\times\vec{m}(x+\Delta x)]\cdot\vec{m}(x-\Delta x) \neq 0$. Assuming linear modulation of the helix, $\phi(x)=qx$, this relation becomes
\begin{align}
 0&\neq 2\sin^2\alpha\cos\alpha\sin(q\Delta z)\left[1-\cos(q\Delta z)\right]\approx\\
 &\approx(q\Delta z)^3\sin^2\alpha\cos\alpha = 2\Delta\varphi_s(q\Delta x)^2(1+\cos\alpha)\cos\alpha,
\end{align}
where $\Delta\varphi_s$ is the spin geometric phase introduced in Eq.~\eqref{eqn:phi_s}. Consequently, intrinsically coplanar states are obtained in the cases of (a) $\alpha=0,\pi$ or $q=0$ (uniform ferromagnetic state), (b) $\alpha=\pi/2$ (helimagnetic state with in-plane magnetization), and (c) $\Delta\varphi_s=0$. Therefore, similarly to the case of ferromagnetic trilayers, the noncoplanarity condition is closely related to a nonvanishing spin geometric phase. From now on, we assume that $\Delta\varphi'=\Delta\varphi+\Delta\varphi_s$ is the total geometric phase, assuming the system allows for both. Since the physics of the effect is not affected by the choice of $\Delta\varphi'$, we do not consider separately the cases of $\Delta\varphi'=\Delta\varphi$ and $\Delta\varphi'=\Delta\varphi_s$.

The inversion-symmetric magnetization profile, which leads to the vanishing of the total geometric phase, $\Delta\varphi'=0$, yields [see Eq.~\eqref{JE}]
\begin{align}
    E_J(\Delta\chi) = -\frac{\hbar}{2}\sum_{\mu,\nu=-\infty}^\infty I_{\mu,\nu}\cos[(\mu+\nu)\Delta\chi].
\end{align}
Apparently, the Josephson energy is an even function of the superconducting difference, $E_J(\Delta\chi)=E_J(-\Delta\chi)$, implying $I_\mathrm{ch/sp}(\Delta\chi)=-I_\mathrm{ch/sp}(-\Delta\chi)$. Note that the spin current in this case is computed by first taking a derivative with respect to $\Delta\varphi'$, then setting it to zero. Apparently, both the normal charge and the normal spin Josephson effects appear. Therefore, one of the necessary conditions for the charge and spin JDE is a nonvanishing total geometric phase induced by a noncoplanar spin arrangement, $\Delta\varphi'\neq 0$. However, this condition is not sufficient, as we discuss below.
\subsection{Different band-specific density of states}
The second ingredient, whose role we are interested in, is the density of states of the two spin bands. In the regime of a weak spin polarization, $J\ll E_F$, they can be considered equal. Translating this regime into the language of Eq.~\eqref{JE} means $I_{\mu,\nu}=I_{\nu,\mu}$, which allows us to draw some general conclusions\footnote{Note that in this case, mixed-spin correlations cannot be neglected in short junctions. However, in long junctions, they are strongly suppressed as compared to equal-spin correlations, making our model valid even in the regime of a weak spin polarization.}. Taking into account the identity
\begin{equation}
    I_{\mu,\nu}\cos\psi_{\mu,\nu}+I_{\nu,\mu}\cos\psi_{\nu,\mu} = 2 I_{\mu,\nu}\cos[(\mu+\nu)\Delta\chi]\cos[(\mu-\nu)\Delta\varphi'],
\end{equation}
yields the following expression for the Josephson energy:
\begin{equation}
    E_J=-\hbar\sum_{\mu\geq\nu} I_{\mu,\nu}\cos[(\mu+\nu)\Delta\chi]\cos[(\mu-\nu)\Delta\varphi']= -\hbar\sum_{\mu\geq\nu}\tilde{I}_{\mu,\nu}\cos[(\mu+\nu)\Delta\chi],
\end{equation}
where $\tilde{I}_{\mu,\nu}\equiv \tilde{I}_{\mu,\nu}(\Delta\varphi')=I_{\mu,\nu}\cos[(\mu-\nu)\Delta\varphi']$. Apparently, the Josephson energy is an even function of the superconducting phase difference $\Delta\chi$ implying the normal charge and spin Josephson effects. The charge Josephson CPR is an odd function of the superconducting phase difference, $I_\mathrm{ch}(\Delta\chi) = -I_\mathrm{ch}(-\Delta\chi)$, and the spin Josephson CPR is an even function of the superconducting phase difference, $I_\mathrm{sp}(\Delta\chi) = I_\mathrm{sp}(-\Delta\chi)$. Since it is odd in $\Delta\chi$, the charge CPR displays no diode effect. However, for the spin CPR, which is an even function in $\Delta\chi$ this is not immediately obvious. To show it, we use the definition of the spin diode efficiency, $\eta_\mathrm{sp}$, proposed in Refs.~\cite{sunVoltagetunableSpinSupercurrent2025a,schulzQuantumgeometricSpinCharge2025}, where
\begin{equation}\label{eqn:diode_sp}
     \eta_\mathrm{sp}=\frac{|I_\mathrm{sp}^+|-|I_\mathrm{sp}^-|}{|I_\mathrm{sp}^+|+|I_\mathrm{sp}^-|},
\end{equation}
with $I_\mathrm{sp}^\pm \equiv I_\mathrm{sp}(\Delta\chi_\mathrm{ch}^\pm)$. Here, $\Delta\chi^\pm$ are the superconducting phase differences at which the positive or negative critical \textit{charge current} is reached for fixed $\Delta\varphi'$, i.e, $\Delta\chi_\mathrm{ch}^+=\mathrm{argmax}_{\Delta\chi}(I_\mathrm{ch})$ and $\Delta\chi_\mathrm{ch}^-=\mathrm{argmin}_{\Delta\chi}(I_\mathrm{ch})$. Since $I_\mathrm{ch}(\Delta\chi)$ is an odd function, it implies $\Delta\chi_\mathrm{ch}^+=-\Delta\chi_\mathrm{ch}^-$. Consequently, $I_\mathrm{sp}^+\equiv I_\mathrm{sp}(\Delta\chi_\mathrm{ch}^+)=I_\mathrm{sp}(-\Delta\chi_\mathrm{ch}^+)=I_\mathrm{sp}(\Delta\chi_\mathrm{ch}^-)\equiv I_\mathrm{sp}^-$, which implies no spin diode effect, $\eta_\mathrm{sp}=0$. Therefore, the second necessary condition for the appearance of the Josephson diode effect is a difference in the densities of states for the two spin bands of the magnetic material, present only in the regime of strong spin polarization. 

\subsection{Restriction to geometric phases and Fourier coefficients}
In the preceding subsections, we have formulated two necessary conditions for the occurrence of the Josephson diode effect across strongly spin-polarized materials: (i) a noncoplanar spin arrangement that leads to a finite geometric phase and (ii) different densities of states of the two spin bands. Both conditions are deduced from the requirement of the absence of a phase-inversion center in the CPRs. However, these two conditions are not sufficient per se, and here we discuss the restriction on the particular values that they can take. Let us first discuss the restrictions to the geometric phase $\Delta\varphi'$, rewriting the Josephson energy~\eqref{JE} as
\begin{equation}
    E_J=-\frac{\hbar}{2}\sum_{\mu,\nu=-\infty}^\infty I_{\mu,\nu}\cos[(\mu+\nu)(\Delta\chi-\chi_0)+\mu(\chi_0-\Delta\varphi')+\nu(\chi_0+\Delta\varphi')].
\end{equation}
Provided $I_{\mu,\nu}$ is nonzero for at least one $\mu\ne 0$ and at least one $\nu \ne 0$, the above equation has a phase-inversion center if $\chi_0-\Delta\varphi'=m\pi$ and $\chi_0+\Delta\varphi'=n\pi\implies \Delta\varphi=k\pi/2$, where $k=n-m$ and $m,n\in\mathbb{Z}$. Note that $k\in 2\mathbb{Z}$ is a special case that refers to the coplanar spin arrangement discussed earlier. The Josephson energy takes the form
\begin{align}
    E_J\qty[\Delta\chi,\Delta\varphi'=\frac{k\pi}{2}]&=-\frac{\hbar}{2}\sum_{\mu,\nu}I_{\mu,\nu}\cos\qty[(\mu+\nu)\qty(\Delta\chi-\frac{k\pi}{2})+\nu k\pi]=\nonumber\\
    &\,=-\frac{\hbar}{2}\sum_{\mu,\nu}(-1)^{\nu k}I_{\mu,\nu}\cos\qty[(\mu+\nu)\qty(\Delta\chi-\frac{k\pi}{2})],
\end{align}
which clearly refers to a $\chi_0$-junction. However, due to the phase-inversion center of the CPR at $\chi_0=k\pi/2$, the Josephson diode effect is absent.  

Let us now consider the restriction on the Fourier coefficients. Our model can be used to show that the JDE is possible only if both spin bands participate in transport. If in the discussion above, $I_{\mu,\nu}=0$ for all $\mu $, then for a phase-inversion center to exist only the condition, $\chi_0=- \Delta \varphi' + n\pi $, needs to be met, i.e. a $\chi_0$-junction is realized for all $\Delta \varphi'$. A similar statement holds for the case that $I_{\mu,\nu}=0$ for all $\nu $.
Therefore, the JDE is absent in a fully spin-polarized ferromagnet, also known as a half-metal. To show this explicitly, we consider our phenomenological model in which the half-metallic regime is obtained simply by switching off one spin channel, i.e., keeping only terms with $\mu=0$ or $\nu=0$. For illustration, let us consider the case of only $\uparrow\uparrow$- mediated transport, i.e., $\nu=0$, for which the Josephson energy reads
\begin{equation}
    E_J(\Delta\chi,\Delta\varphi')=-\frac{\hbar}{2}\sum_{\nu=-\infty}^\infty I_{\mu,0}\cos[\mu(\Delta\chi-\Delta\varphi')].
\end{equation}
Apparently, such a situation corresponds to a $\chi_0$-junction whose CPR possesses a phase-inversion center at $\chi_0=\Delta\varphi'$, and the JDE is absent. Therefore, the effect is only possible if both spin bands contribute to transport. In what follows, we provide a minimal model which summarizes all the necessary conditions mentioned above. 
\subsection{Minimal model}
\begin{figure}[t!]
    \centering
    \includegraphics[width=1\linewidth]{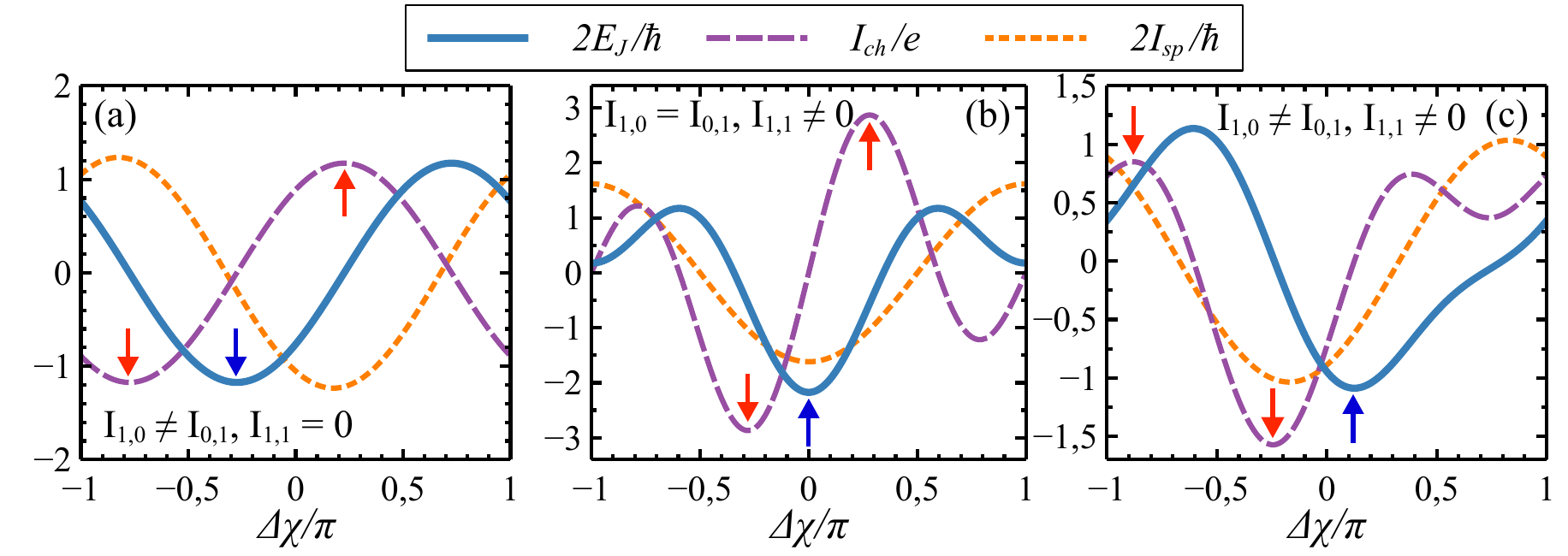}
    \caption{The Josephson energy (blue), the charge current (violet), and the spin current (orange) computed from the minimal model of Eq.~\eqref{eqn:Ej_minimal} for (a) a strongly spin-polarized junction in the tunneling limit $I_{1,1}=0$, (b) a weakly spin-polarized highly transmissive junction $I_{1,0}=I_{0,1}$ and $ I_{1,1}\neq 0$, and (c) a strongly spin-polarized highly transmissive junction, $I_{1,0}\neq I_{0,1}$ and $I_{1,1}\neq 0$. In all panels, $\Delta\varphi'\neq k\pi/2$.}
    \label{fig:model}
\end{figure}

 As shown in Refs.~\cite{schulzQuantumgeometricSpinCharge2025,schulzTheoryQuantumgeometricCharge2025a,nikolicSpinresolvedJosephsonDiode2025} to resemble Josephson CPRs, it is typically sufficient to keep just a few terms in expansion~\eqref{JE}. Naturally, the first guess would be to account for contributions that involve the transfer of one Cooper pair, namely, keeping just the coefficients $I_{0,1}$ and $I_{1,0}$ in Eq.~\eqref{JE}. This situation corresponds to the tunneling limit, and the Josephson energy takes the form
 \begin{equation}
 \begin{split}
          E_J(\Delta\chi,\Delta\varphi')&\approx-\frac{\hbar}{2}[I_{1,0}\cos(\Delta\chi-\Delta\varphi')+I_{0,1}\cos(\Delta\chi+\Delta\varphi')]=\\
     &=-\frac{\hbar}{2}I_c(\Delta\varphi')\cos(\Delta\chi-\chi_0),
 \end{split}
 \end{equation}
 where $I_c=\sqrt{I_{1,0}^2+I_{0,1}^2+2I_{1,0}I_{0,1}\cos(2\Delta\varphi')}$ and $\chi_0 = \frac{I_{1,0}-I_{0,1}}{I_{1,0}+I_{0,1}}\tan(\Delta\varphi')$. Therefore, in the tunneling limit, only the anomalous Josephson effect ($\chi_0$-junction) is possible, and it vanishes in the case of a coplanar spin texture, $\Delta\varphi'=0$, and weak spin polarization, $I_{1,0}=I_{0,1}$, which is in accordance with our preceding discussion. The tunneling limit is illustrated in Fig.~\ref{fig:model}(a), where the minimum of the Josephson energy is located at $\chi_0\neq k\pi, k\in\mathbb{Z}$ (blue arrow) and a $\chi_0$-junction is realized (red arrows). 

 Therefore, to formulate a minimal model that displays the JDE, we need to account for higher-order processes that involve at least two Cooper pairs~\cite{schulzTheoryQuantumgeometricCharge2025a}. The corresponding Josephson energy~\eqref{JE}, therefore, reads
 \begin{align}
     E_J(\Delta\chi,\Delta\varphi')\approx &-\frac{\hbar}{2}\big[I_{1,0}\cos(\Delta\chi-\Delta\varphi')+I_{0,1}\cos(\Delta\chi+\Delta\varphi') +\\
     &+I_{1,1}\cos(2\Delta\chi)+I_{2,0}\cos(2\Delta\chi-2\Delta\varphi') + I_{0,2}\cos(2\Delta\chi+2\Delta\varphi')\big].\nonumber
 \end{align}
However, it turns out that in highly transmissive junctions, among higher order contributions, the so-called crossed-pair transmission processes are typically the most dominant ones, i.e., $I_{1,1}>I_{2,0},I_{0,2}$~\cite{schulzPurelyEvenHarmonic2025}. Therefore, we further simplify the above equation:
\begin{equation}\label{eqn:Ej_minimal}
     E_J(\Delta\chi,\Delta\varphi')\approx -\frac{\hbar}{2}[I_{1,0}\cos(\Delta\chi-\Delta\varphi')+I_{0,1}\cos(\Delta\chi+\Delta\varphi')+I_{1,1}\cos(2\Delta\chi)].
\end{equation}
The corresponding charge and spin currents read, respectively,
\begin{align}
    \label{eqn:Ich_minimal}
    I_\mathrm{ch}(\Delta\chi,\Delta\varphi')&\approx e[I_{1,0}\sin(\Delta\chi-\Delta\varphi')+I_{0,1}\sin(\Delta\chi+\Delta\varphi')+2I_{1,1}\sin(2\Delta\chi)],\\
    \label{eqn:Isp_minimal}
    I_\mathrm{sp}(\Delta\chi,\Delta\varphi')&\approx \frac{\hbar}{2}[I_{1,0}\sin(\Delta\chi-\Delta\varphi')-I_{0,1}\sin(\Delta\chi+\Delta\varphi')].
\end{align}
Note that the crossed-pair processes are not mandatory for the charge JDE. Other higher-order processes could also yield the effect. However, as mentioned above, these processes typically contribute less than the crossed-pair ones~\cite{schulzTheoryQuantumgeometricCharge2025a}. Equations~\eqref{eqn:Ej_minimal}-\eqref{eqn:Isp_minimal} represent the minimal model that describes the Josephson diode effect across strongly spin-polarized materials, and it can be used for a qualitative description of the effect.

As already mentioned, Fig.~\ref{fig:model}(a) shows the tunneling limit with higher-order processes being strongly suppressed, $I_{1,1}=0$. An anomalous $\chi_0$-junction with no diode effect is realized. Figure~\ref{fig:model}(b) shows the case of a highly transmissive junction which allows for higher order processes, $I_{1,1} \neq 0$, however, in a weakly spin-polarized regime, $I_{1,0}=I_{0,1}$. The minimum of the Josephson energy (blue arrow) is at $\Delta\chi=0$, implying a normal $0$-junction. The minimum and maximum of the charge current, denoted by the red arrows, have the same absolute value and are achieved for the same absolute value of $\Delta\chi$. Finally, Fig.~\ref{fig:model}(c) shows a highly transmissive, $I_{1,1}\neq 0$, strongly spin-polarized junction, $I_{1,0}\neq I_{0,1}$, which exhibits both the charge and the spin Josephson diode effect. In particular, the Josephson energy is minimum (blue arrow) for $\Delta\chi\neq k\pi$ ($k\in\mathbb{Z}$) possessing no phase-inversion center. Consequently, the charge current has no phase-inversion center either, and the absolute values of the maximum and the minimum current differ (red arrows).

\begin{figure}[t!]
    \centering
    \includegraphics[width=0.5\linewidth]{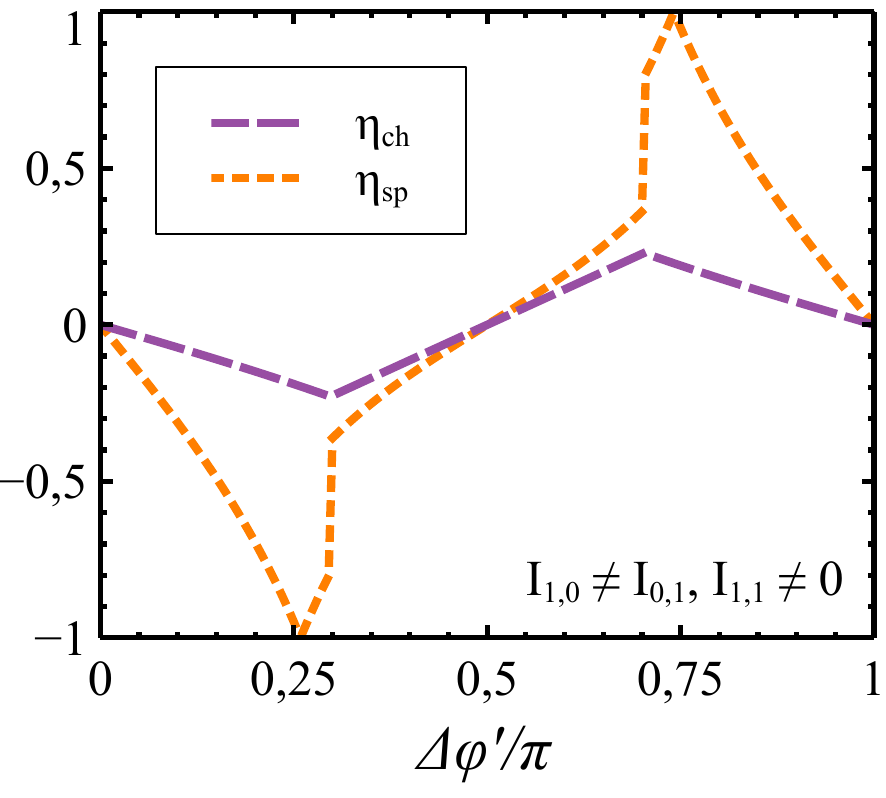}
    \caption{The charge ($\eta_\mathrm{ch}$; dashed violet) and spin ($\eta_\mathrm{sp}$; dotted orange) diode efficiencies as functions of the geometric phase $\Delta\varphi'$ in a highly transmissive junction, $I_{1,1}\neq 0$, involving a strongly spin-polarized magnetic material, $I_{1,0}\neq I_{0,1}$.}
    \label{fig:diode}
\end{figure}

The minimal model of Eq.~\eqref{eqn:Ej_minimal} can be used for calculating the diode efficiencies shown in Fig.~\ref{fig:diode}. The dashed violet line shows the charge diode efficiency $\eta_\mathrm{ch}$ as a function of the geometric phase $\Delta\varphi'$. Since the model is qualitative and the existence of the effect is not sensitive to particular values of the coefficients as long as $I_{1,0}\neq I_{0,1}$ and $I_{1,1}\neq 0$, we do not comment on them here. The corresponding spin diode efficiency $\eta_\mathrm{sp}$ (see Eq.~\eqref{eqn:diode_sp} and the text below) is shown by the doted orange line. Remarkably, even a simple model such as Eq.~\eqref{eqn:Ej_minimal} can yield significant charge diode effect and a perfect spin diode effect.

\section{Summary}
In summary, we have presented a set of conditions necessary for the appearance of charge and spin Josephson diode effects across a strongly spin-polarized magnetic material placed between two conventional spin-singlet superconductors. In particular, we have considered two distinct cases: (i) a ferromagnetic trilayer consisting of a central metallic ferromagnet coupled to two thin satellite ferromagnetic insulators and (ii) a metallic conical magnet. 

Upon defining nonlocal geometric phases, $\Delta\varphi'$, induced by a noncoplanar spin texture in magnetic materials, we have examined their role in the Josephson effect. Due to their U(1) nature, we have shown that they enter the Josephson current-phase relation (CPR) similarly to the superconducting phase difference, $\Delta\chi$. As a consequence, the CPR is modified, giving rise to both the anomalous and the Josephson diode effects. Even though the former is allowed in a CPR that possesses a phase-inversion center with respect to $\Delta\chi$, the latter is absent in this case. 

We have shown that the phase-inversion center in the CPR is absent under the following conditions. First, the noncoplanarity of the spin texture that breaks the spatial inversion symmetry, giving rise to a nonvanishing $\Delta\varphi'$, is required. Second, different densities of states for the two spin bands achieved by strong spin polarization are demanded. Third, both spin bands must contribute to transport, i.e., the effect is absent in the case of a half-metal. Fourth and last, higher harmonics in the CPR are demanded, i.e., the effect is absent in the tunneling limit. However, higher harmonics are not sufficient per se; namely, the CPR still must have no phase-inversion center. This situation is achieved by restricting the quantum geometric phases to values $\Delta\varphi'\neq k\pi/2, k\in\mathbb{Z}$. All these conditions can be naturally met in a strongly spin-polarized ferromagnetic (but not half-metallic) Josephson junction with a non-coplanar spin texture and appreciable interface transmissions.

We illustrate our theory by formulating a simple model involving up to two Cooper pairs transferred across the junction. As a result, we have shown that even a simple model incorporating the aforementioned points can feature a significant charge and a perfect spin diode effect. 

Finally, we stress that considering a ferromagnetic trilayer does not require the use of outer ferromagnetic insulators. The physics of the effect presented in this work will remain intact if a metallic ferromagnetic trilayer is used, provided that the satellite ferromagnets are sufficiently thin, allowing all three triplet components to penetrate appreciably. The important issue is the misalignment of the exchange fields, which leads to triplet rotation and allows equal-spin pairs to enter the central ferromagnet, and this mechanism works for both metallic and insulating ferromagnetic interface layers.  

\bibliography{sn-bibliography}

\end{document}